\documentclass{emulateapj}
\usepackage{natbib}
\usepackage[usenames,dvipsnames]{color}
\usepackage{amsmath}

\shorttitle{IMPLICATIONS FOR BSS FORMATION WITH HST}
\shortauthors{GOSNELL ET AL.}

\begin{document}
\title{Implications for the Formation of Blue Straggler Stars from HST Ultraviolet Observations of NGC 188\footnote{WIYN Open Cluster Survey Paper LXX.}}

\author{Natalie M. Gosnell\altaffilmark{1,2}, Robert D. Mathieu\altaffilmark{2}, Aaron M. Geller\altaffilmark{3,4}, Alison Sills\altaffilmark{5}, Nathan Leigh\altaffilmark{6,7}, Christian Knigge\altaffilmark{8}}
\email{gosnell@astro.as.utexas.edu}
\altaffiltext{1}{Department of Astronomy, The University of Texas at Austin, 2515 Speedway, Stop C1400, Austin, TX  78712-1205}
\altaffiltext{2}{Department of Astronomy, University of Wisconsin - Madison, 475 N. Charter Street, Madison, WI 53706}
\altaffiltext{3}{Center for Interdisciplinary Exploration and Research in Astrophysics (CIERA) and Department of Physics and Astronomy, Northwestern University, 2145 Sheridan Rd, Evanston, IL  60208}
\altaffiltext{4}{Department of Astronomy and Astrophysics, University of Chicago, 5640 S. Ellis Avenue, Chicago, IL 60637}
\altaffiltext{5}{Department of Physics and Astronomy, McMaster University, 1280 Main St. W, Hamilton, ON  L8S 4M1, Canada}
\altaffiltext{6}{Department of Physics, University of Alberta, CCIS 4-183, Edmonton, AB  T6G 2E1, Canada}
\altaffiltext{7}{Department of Astrophysics, American Museum of Natural History, Central Park West and 79th Street, New York, NY 10024}
\altaffiltext{8}{School of Physics and Astronomy, University of Southampton, Highfield, Southampton, SO17 IBJ, UK}

\begin{abstract}
We present results of a \textit{Hubble Space Telescope} far-ultraviolet (FUV) survey 
searching for white dwarf (WD) companions to blue straggler stars (BSSs) in  
open cluster NGC 188. The majority of NGC 188 BSSs (15 of 21) are single-lined 
binaries with properties suggestive of mass-transfer formation via Roche lobe overflow, 
specifically through an asymptotic giant branch star transferring mass to a main 
sequence secondary, yielding a BSS binary with a WD companion. In NGC 188, a 
BSS formed by this mechanism within the past 400 Myr will have a WD companion 
hot and luminous enough to be directly detected as a FUV photometric excess with 
\textit{HST}. Comparing expected BSS FUV emission to observed photometry 
reveals four BSSs with WD companions above 12,000 K (younger than 250 Myr) and 
three WD companions with temperatures between 11,000--12,000 K. These BSS+WD 
binaries all formed through recent mass transfer. The location of the young BSSs in 
an optical color-magnitude diagram (CMD) indicates that distance from the zero-age 
main sequence does not necessarily correlate with BSS age. There is no clear CMD 
separation between mass transfer-formed BSSs and those likely formed through other 
mechanisms, such as collisions. The seven detected WD companions place a lower 
limit on the mass-transfer formation frequency of 33\%. We consider other possible 
formation mechanisms by comparing properties of the BSS population to theoretical 
predictions. We conclude that 14 BSS binaries likely formed from mass transfer, 
resulting in an inferred mass-transfer formation frequency of approximately 67\%. 
\end{abstract}

\section{INTRODUCTION}
Open clusters are ideal environments for studying stellar populations. The ability to isolate a single-age stellar population from the field using 
radial-velocity (RV) and proper-motion measurements not only informs our understanding of normal single-star evolution, but also highlights 
those stars whose evolutionary paths deviate from single-star expectations. One such group of objects is blue 
straggler stars (BSSs). Traditionally defined to be star cluster members that are brighter and bluer than the corresponding main 
sequence turnoff \citep{Sandage53}, the definition extends to stars that appear to be too young given the cluster age, 
such as stars below the turnoff but blueward of the main sequence \citep{Mathieu09}. BSSs are not anomalous 
objects; they are found in open clusters \citep[e.g.,][]{Johnson55,Burbidge58,Leonard96,Sandquist05,Talamantes10}, globular clusters 
\citep[e.g.,][]{Sandage53,Ferraro99,Piotto04,Leigh07,Knigge09,Santana13}, the Galactic field \citep{Preston00,Carney01}, and 
dwarf galaxies \citep{Momany07,Mapelli07}. The BSS population of open cluster NGC 188, in addition to a few sub-subgiants and yellow giants, 
comprises approximately 25\% of the evolved cluster population \citep[][and references therein]{Mathieu14}. Thus, 
understanding the creation and subsequent evolution of BSSs 
and other non-standard stellar products is 
fundamental to completing the full picture of stellar evolution at large.

BSSs were first discovered about 60 years ago \citep{Sandage53,Johnson55}, and understanding their origin continues to be an active field 
of research. A common theme is that BSS progenitors must gain additional mass in order to appear above the main sequence turnoff, but 
theories differ on how the mass is acquired. 
There are currently three primary 
theories for binary BSS formation: stellar collisions during dynamical interactions of binaries with single stars or other
binaries \citep[e.g.,][]{Leonard89,Ferraro95,Mapelli06,Leigh11}, the merger of an inner binary in a triple system via 
the Kozai mechanism \citep{Perets09,Naoz14}, and Roche lobe overflow mass transfer while 
the original (progenitor) primary star is on the red giant branch (RGB, Case B) or asymptotic 
giant branch \citep[AGB, Case C; e.g.,][]{McCrea64,Ferraro01,Ferraro06,Chen08,Leigh11b,Leigh13,Gosnell14}. 
Although not a likely pathway for binary BSSs, mass transfer while both stars 
are on the main sequence (Case A) can create single BSSs \citep{Webbink76,Lombardi02}.

Broadly, all these formation scenarios create BSSs similar to those observed in star clusters, but the details of the BSS population 
will vary. Collisions can create single BSSs or binary BSSs that retain a companion from the dynamical 
encounter \citep[e.g.,][]{Hurley05,Leigh11,Leigh12}. 
(In this work, a collision product is a star resulting from a collision, coalescence, or merger of two stars during a dynamical encounter.)
BSSs formed through Case A mass transfer typically lack a binary companion \citep{Webbink76,Lombardi02}, although
\citet{Tian06} model the formation of BSSs from Case A mass transfer that result in binaries with periods less than 10 days.  
The Kozai mechanism will create a binary BSS
whose secondary is the original tertiary member of the system \citep{Perets09}. A BSS formed from 
Roche lobe overflow will also be in a binary, but the secondary star will be the core
of the progenitor primary star, observed as a carbon/oxygen (CO) or helium white dwarf \citep[WD;][]{Geller11,Geller13,Gosnell14}. 

The number of BSSs observed in a star cluster is determined by a combination of the formation rate for a given mechanism and the 
corresponding BSS lifetime.  
$N$-body models are capable of predicting the number of BSSs in cluster environments created through each 
formation method \citep{Hurley05,Geller13}. However, empirical determination of BSS formation 
for a given population is necessary in order to make appropriate comparisons to $N$-body model results.

Observational identification of the particular formation mechanisms for an entire population of BSSs has proved elusive, 
although formation pathways for several individual BSSs have been determined in globular clusters. 
\citet{Knigge08} find a globular cluster BSS+WD binary that perhaps formed through mass transfer (although it may 
also be the result of a dynamical exchange) 
and \citet{Rozyczka13} find a globular cluster BSS currently undergoing mass transfer.
In 47 Tuc, \citet{Ferraro06} discover five BSSs with depleted CO that is suggestive of a mass transfer history and 
\citet{Knigge06} find a BSS likely formed from at least three stars,  
but (by design) none of these studies provide enough context to learn about formation mechanisms of the
entire BSS population. In globular cluster M30, \citet{Ferraro09} find a trend in 
optical color and magnitude that suggest two BSS sub-populations split between mass transfer- and 
collision-formed products, 
but further observational confirmation, such as detecting BSS binarity, is very difficult 
due to the crowded nature of globular clusters. Finally, the existence of W UMa-type binaries in the BSS region 
of globular clusters \citep[e.g.,][]{Rucinski00} is evidence that mass transfer must be responsible for at 
least a portion of globular cluster BSS populations.

In an attempt to disentangle the problem of multiple formation pathways we focus on the population of BSSs in open 
cluster NGC 188, which presents an excellent environment for studying 
these objects. The cluster is one of the oldest known open clusters in the Galaxy, with 
an age of 7 Gyr \citep{Sarajedini99}. The BSSs, therefore, are not very ``blue,'' 
allowing for accurate RV measurements and binary orbit determinations \citep{Geller08,Geller09}.
NGC 188 is also relatively close \citep[$1770\pm75$ pc;][]{Meibom09} so 
observations from both ground and space are not prohibitively expensive. The WIYN Open Cluster Study 
\citep{Mathieu00} has observed NGC 188 for almost 20 years, obtaining 
photometry \citep{Sarajedini99}, proper motions \citep{Platais03}, RVs \citep{Geller08}, and 
binary orbits \citep{Geller09}, and has expanded into theoretical efforts 
with sophisticated $N$-body models covering the entire cluster lifetime \citep{Geller13}. 
This enormous foundation of data makes the NGC 188 BSS population one of the most well-studied in the Galaxy.

The NGC 188 BSS population includes 21 stars, 20 of which are high-probability 
3-dimensional cluster members \citep{Platais03,Geller08}. 
The single exception, WOCS 4230, is a binary star photometrically classified as a 
BSS but lacks a solved binary orbit due to rapid rotation, 
resulting in an unknown RV-determined cluster membership. (\citealt{Platais03} 
determine a proper motion membership probability for WOCS 4230 of 93\%.)
The majority (80\%) of the BSSs in NGC 188 are spectroscopic binaries \citep{Mathieu09}. Four BSSs 
do not have detected RV variability, meaning they are not in binaries with orbital periods less than $10^{4}$ days \citep{Geller09}. 
They either have companions at very long periods, $10^{4} < P_{\mathrm{orb}} < 10^{6}$ days, or are single stars \citep{Geller12}. 
Binaries with periods beyond $10^{6}$ days are thought to be destroyed through dynamical interactions in the cluster environment. 
The BSS population includes two double-lined (SB2) binaries with very short orbital periods around 5 days and high mass 
ratios of $q=1.0$ and $q=0.68$ \citep{Geller09}. The remaining 15 BSS binaries are single-lined, which we refer to as SB1 systems. 
In Table~\ref{tab:bssinfo} we present the NGC 188 BSS population. We list the 
identification number (WOCS ID), J2000 location, optical photometry and 
color, membership class, and effective temperature. In addition we include the orbital period and eccentricity for binary BSSs 
with orbital solutions. 

\begin{table*}
\begin{center}
\footnotesize
\caption{NGC 188 Blue Straggler Star Population} \label{tab:bssinfo}
 \vspace{3 mm}
\begin{tabular*}{\textwidth}{@{\extracolsep{\fill}} lccccccc}
ID & Position (J2000) & $V^\mathrm{a}$ & $B-V^\mathrm{a}$ & Class$^\mathrm{b}$ & $T_{\mathrm{eff}}^\mathrm{c}$  & $P_{\mathrm{orb}}$ (days)$^\mathrm{b}$ & Eccentricity$^\mathrm{b}$ \\\hline
\multicolumn{6}{l}{\textit{Non-velocity variable:}}\\
1366  & 00:51:15.06, +85:44:02.02 & 15.851 & 0.620 & SM & $6120\pm120$  & ... & ... \\
4290  & 00:42:06.53, +85:16:47.25 & 14.174 & 0.584 &   SM  & $6280\pm\phantom{0}90$ & ... & ... \\
4306  & 00:42:20.59, +85:15:39.47 & 13.347 & 0.534 &  SM & $6450\pm100$ &  ... & ...  \\
5020  & 00:47:51.46, +85:15:09.09 & 14.000 & 0.502 &  SM & $6750\pm130$ & ... & ... \\   
\multicolumn{6}{l}{\textit{Single-lined binaries:}}\\     
451    & 00:34:47.95, +85:32:27.33 & 13.880 & 0.604 & BM & $6400\pm110$ & $\phantom{0}722\phantom{.00}\pm\phantom{00}4\phantom{.00}$ & $0.34\phantom{0}\pm0.03\phantom{0}$ \\
1888  & 00:54:31.35, +85:32:09.12 & 14.841 & 0.552 &  BM & $6570\pm120$ &  $2240\phantom{.00}\pm\phantom{0}30\phantom{.00}$ & $0.21\phantom{0}\pm0.04\phantom{0}$   \\
2679  & 00:26:44.64, +85:18:35.94 & 15.011 & 0.515 & BM & $6630\pm120$ & $1033\phantom{.00}\pm\phantom{00}8\phantom{.00}$ & $0.07\phantom{0}\pm0.05\phantom{0}$ \\            
4230  & 00:43:23.81, +85:20:32.64 & 15.080 & 0.534 & BU & $6350\pm110$ &  ... & ...\\            
4348  & 00:43:41.47, +85:13:17.28 & 14.681 & 0.466 &  BM & $6750\pm120$ & $1168\phantom{.00}\pm\phantom{00}8\phantom{.00}$ & $0.09\phantom{0}\pm0.05\phantom{0}$ \\            
4540  & 00:45:18.27, +85:19:19.85 & 13.857 & 0.521 & BM & $6590\pm100$  & $3030\phantom{.00}\pm\phantom{0}70\phantom{.00}$ & $0.36\phantom{0}\pm0.07\phantom{0}$ \\             
4581  & 00:45:56.63, +85:17:29.66 & 14.147 & 0.536 &  BM & $6750\pm100$ & $\phantom{0}546.7\phantom{0}\pm\phantom{00}1.6\phantom{0}$ & $0.269\pm0.015$ \\
4589  & 00:46:22.99, +85:17:13.46 & 14.995 & 0.595 & BM & $6240\pm100$ & $\phantom{0}615.2\phantom{0}\pm\phantom{00}1.7\phantom{0}$ & $0.21\phantom{0}\pm0.04\phantom{0}$ \\
4970  & 00:47:13.32, +85:16:39.87 & 14.524 & 0.622 &  BM & $6170\pm\phantom{0}90$ & $1002.76\pm\phantom{00}2.4\phantom{0}$ & $0.095\pm0.013$ \\
5325  & 00:49:36.86, +85:16:38.85 & 14.978 & 0.605 &   BM  & $6060\pm100$ & $1772\phantom{.00}\pm\phantom{0}12\phantom{.00}$ & $0.77\phantom{0}\pm0.03\phantom{0}$    \\
5350  & 00:49:03.26, +85:15:25.19 & 13.453 & 0.495 & BM & $6720\pm120$ & $\phantom{0}690\phantom{.00}\pm \phantom{00}3\phantom{.00}$ & $0.07\phantom{0}\pm0.03\phantom{0}$  \\         
5379  & 00:50:10.79, +85:14:38.08 & 15.372 & 0.570 & BM & $6400\pm120$ & $\phantom{0}120.21\pm \phantom{00}0.04$ & $0.24\phantom{0}\pm0.03\phantom{0}$ \\          
5434  & 00:48:54.45, +85:12:36.71 & 14.687 & 0.526 &   BM   & $6550\pm120$ & $1277\phantom{.00}\pm \phantom{00}9\phantom{.00}$ & $0.551\pm0.018$   \\
5671  & 00:52:25.78, +85:15:27.88 & 13.600 & 0.659 &    BM & $6130\pm\phantom{0}90$ & $1423\phantom{.00}\pm \phantom{00}7\phantom{.00}$ & $0.286\pm0.018$  \\
8104  & 00:40:15.45, +85:03:48.49 & 15.842 & 0.616 &   BM   & $6170\pm130$ & $2140\phantom{.00}\pm110\phantom{.00}$ & $0.20\phantom{0}\pm0.07\phantom{0}$   \\ 
\multicolumn{6}{l}{\textit{Double-lined binaries:}} \\
5078  & 00:47:11.69, +85:13:31.53 & 14.465 & 0.601 & BM & ... & $4.78303\pm0.00012$ & $0.121\pm0.0006$ \\
7782  & 00:35:46.23, +84:57:14.35 & 14.379 & 0.486 & BM & ... & $5.32770\pm0.00005$ & $0.013\pm0.006\phantom{0}$ \\
\hline
\multicolumn{8}{l}{$^{\mathrm{a}}$\citet{Sarajedini99}.}\\
\multicolumn{8}{l}{$^{\mathrm{b}}$\citet{Geller09}. Classification classes are SM: Single Member, BM: Binary Member, and BU: Binary }\\
\multicolumn{8}{l}{$\phantom{^{\mathrm{b}}}$proper motion member, but no orbit or RV membership.}\\
\multicolumn{8}{l}{$^{\mathrm{c}}$Calculated using 4- or 5-band color information following \citet{Ramirez05}, see Section~\ref{sec:bsmodel}.}
\end{tabular*}
\end{center}
\end{table*}

The binary parameters of the SB1 systems are striking. All but one of the SB1 BSSs have 
periods within a factor of 2--3 of 1000 days, compared to the main sequence period distribution that extends from a few days up to several 
thousand days \citep{Mathieu09}.  
The statistical secondary mass distribution has a sharp peak at 0.5 $M_{\odot}$ that is distinct from the 
mass distribution expected for main sequence secondary stars at the 99\% confidence level.  This suggests the companions 
are CO WDs, whose masses are typically around 0.5 $M_{\odot}$ \citep{Geller11}. 
These binary parameters are consistent with the BSSs forming through Case C mass transfer \citep{Chen08}. 

We set out to observationally detect these potential WD companions with a far-ultraviolet (FUV) photometric campaign using 
the \textit{Hubble Space Telescope} (\textit{HST}). 
The first results of this study are presented in \citet{Gosnell14}, where we identify three BSSs with hot WD companions and present specific formation scenarios for 
each system. The shorter period of WOCS 5379 of 120 days opens the 
possibility of Case B rather than Case C mass transfer, which 
is also explored in \citet{Gosnell14}. 

In this paper we present the results of the \textit{HST} study for the remaining BSS population in NGC 188 and discuss the implications for the frequency of 
different BSS formation mechanisms. We present our observations in Section~\ref{sec:obs}, 
our photometric analysis of the BSS population and detection of WD companions in Section~\ref{sec:analysis}, discuss the 
frequency of mass-transfer 
formation and implications for modeling efforts in 
Section~\ref{sec:bspop}, and summarize our study in Section~\ref{sec:summary}.

\section{OBSERVATIONS}
\label{sec:obs}
Our observational design is described in \citet{Gosnell14}, but the salient details are repeated here 
for clarity. 19 of the 21 BSSs in NGC 188 were observed using the \textit{Hubble Space Telescope} Advanced Camera for Surveys (ACS)  
in the Solar Blind Channel (SBC). The observations occurred in 19 separate 2-orbit visits between 
2012 October 3 and 2012 November 23 (GO: 12492, PI: Mathieu). We did not include the two SB2 BSSs in this study, 
as their high mass ratios and optical detection of the secondary stars indicate the companions 
are not WDs \citep{Geller09,Mathieu09}. An additional star, WOCS 1947, was included in the \textit{HST} study but  
\citet{Mathieu14} later found it to be a red giant star. A $V$-band photometric error in the 
literature caused the erroneous categorization of this source. 
We return to WOCS 1947 in Section~\ref{sec:redleak}. 

Each BSS was observed in F140LP for 2040~s, F150LP for 2380~s, and F165LP for 1564~s. The total exposure times 
are the sum of four shorter exposures using the standard ACS/SBC 4-point box dither pattern. We also utilize derived
narrow bandpasses to better isolate the bluest FUV flux. We exploit the nested nature of the long-pass filters by 
differencing F140LP and F150LP to create F140N, and differencing F150LP and F165LP to create F150N \citep{Dieball05, Gosnell14}. 

The SBC field of view is very small at 25\arcsec$\times$25\arcsec. In all but one case the BSS target is the 
only star visible in the SBC image; therefore the images do not suffer from source confusion or blending.

\subsection{ACS/SBC Aperture Photometry}
Aperture photometry is carried out on redrizzled images with a pixel scale of 0\farcs025 pixel$^{-1}$. We extract 
count rates using an aperture radius of 6 pixels, or 0\farcs15, using the IRAF package \textsc{daophot}. 
We calculate encircled energy fractions for this aperture size using modeled Tiny Tim point sources\footnote{\texttt{http://www.stsci.edu/hst/observatory/focus/TinyTim}}. 
We find corrected count rates by dividing the measured count rate by the encircled energy corrections of 0.83, 0.84, and 0.85 
for F140LP, F150LP, and F165LP, respectively. 

\subsubsection{ACS Red Leak}
\label{sec:redleak}	
The ACS/SBC has a known red leak beyond 2000\AA\ \citep{Boffi08}. Recent work has also shown that 
the red leak model included in \textsc{synphot} is no 
longer accurate \citep{Feldman10}. Although the SBC detector temperature rises with continued use, 
we do not find any systematic correlation between flux and detector temperature across the four dithered 
exposures for each target.

In the absence of identifying a red leak trend we use the serendipitous observation of the red giant WOCS 
1947 to account for red leak contamination. As a red giant \citep[$V=12.54$, $B-V=1.29$;][]{apass}, we assume 
all of the SBC flux for WOCS 1947 is due to the red leak. We estimate the red leak contribution of each BSS by 
scaling the detected red leak flux of WOCS 1947 in each long-pass filter by $V$-band luminosity. We then subtract off the ``extra'' red leak flux. 
This results in count rates that, to the best of current capabilities, are red leak-free. The mean red leak correction in F150LP is 0.15 mag.

The detected red leak flux for WOCS 1947 is about twice the predicted flux from current \textsc{synphot} throughput curves.
Modeling WOCS 1947 with an ATLAS9 spectrum \citep{atlas9} with an effective temperature of 4575 K and surface gravity of log $(g)=1.5$, 
the \textsc{synphot} task \textsc{calcphot} 
calculates a count rate of 0.12 counts s$^{-1}$ in all three long-pass filters. This is in comparison to the measured count rates of $0.212\pm0.009$, 
$0.279\pm0.009$, and $0.23\pm0.01$ counts s$^{-1}$ for F140LP, F150LP, and F165LP, respectively. The resulting magnitude difference is 
0.9 STMAG in F150LP ($M_{\textsc{synphot}}=22.1$, $M_{\mathrm{obs}}=21.2$). These observations also suggest that the red leak is not 
constant across the long-pass filters. The measured difference of the observed red leak compared to \textsc{synphot} is slightly less than 
the difference seen by \citet{Feldman10}. Using the SBC PR130L prism, they find the 
observed red leak throughput to be 2.5 times that of the \textsc{synphot} throughput curve.  

\subsubsection{Magnitude Calculation}
Encircled energy-corrected and red leak-subtracted count rates are used to calculate instrumental magnitudes in the STMAG system. 
We convert the count rates into fluxes (erg cm$^{-2}$ s$^{-1}$ \AA$^{-1}$) using the 
PHOTFLAM conversion factors for F140LP, F150LP, and F165LP provided in the ACS Data Handbook \citep{acshandbook}. 
We find the count rates in F140N and F150N by differencing the corrected count rates of the long-pass filters, where 
F140N = (F140LP $-$ F150LP) and F150N = (F150LP $-$ F165LP). We calculate the flux conversion factors for F140N and F150N 
using \textsc{synphot} and the F140N and F150N flux errors by combining the flux errors of the long-pass filters in quadrature, 
scaled by the different exposure times. We calculate instrumental magnitudes in all bandpasses using 
$\mathrm{STMAG} = -2.5\,\mathrm{log}_{10}(\mathrm{flux})-21.1.$ 
The photometric errors are dominated by Poisson noise 
due to the low background of the ACS/SBC MAMA detector, although the signal is high enough that the error distributions 
are approximately symmetric for all but the faintest measurements in F140N and F150N. 

The FUV information for the 19 BSS in this study are presented in Table~\ref{tab:bssphot} with the WOCS ID, 
NUV magnitude, and 5-band FUV magnitudes (F140LP, F150LP, F165LP, F140N, and F150N).  A F140N magnitude is 
only reported for those sources with a higher count rate in F140LP than F150LP.  Magnitudes given in italics are less 
than 3$\sigma$ detections.

\begin{table*}
\footnotesize
\caption{\textit{HST} FUV Photometry of NGC 188 BSS Population} \label{tab:bssphot}
\begin{tabular*}{\textwidth}{c @{\extracolsep{\fill}} cccccc}
WOCS ID & NUV$^\mathrm{a}$ & F165LP & F150LP & F140LP & F150N$^\mathrm{b}$ & F140N$^\mathrm{b}$ \\\hline
451    & $18.38\pm0.04$ & $18.94\pm0.04$ & $20.06\pm0.03$ & $20.54\pm0.03$ & $22.2^{+0.5}_{-0.8}$    & $\mathit{23.2^{+0.7}_{-2.0}}$  \\
1366  &   $20.6\phantom{0}\pm0.2\phantom{0}$     & $20.63\pm0.07$ & $21.83\pm0.05$ & $22.62\pm0.06$ & $\mathit{25.5^{+1.7}_{-1.7}}$    &  ...            \\
1888  & $19.39\pm0.07$ & $19.78\pm0.05$ & $20.72\pm0.04$ & $21.37\pm0.04$ & $21.9^{+0.3}_{-0.4}$    &  ...           \\
2679  & $19.07\pm0.06$ & $19.22\pm0.04$ & $20.21\pm0.03$ & $20.67\pm0.03$ & $21.5^{+0.3}_{-0.3}$    &  $\mathit{22.7^{+0.4}_{-0.8}}$ \\            
4230  & $19.49\pm0.07$ & $19.38\pm0.04$ & $20.33\pm0.03$ & $20.77\pm0.03$ & $21.5^{+0.2}_{-0.3}$    &  $\mathit{22.6^{+0.4}_{-0.6}}$  \\            
4290  & $19.02\pm0.06$ & $19.23\pm0.04$ & $20.32\pm0.03$ & $20.90\pm0.03$ & $22.3^{+0.4}_{-0.7}$    &  ...           \\
4306  & $18.00\pm0.04$ & $18.23\pm0.03$ & $19.29\pm0.02$ & $19.81\pm0.02$ & $20.9^{+0.2}_{-0.3}$    &  ...          \\
4348  & $18.59\pm0.05$ & $18.64\pm0.03$ & $19.55\pm0.02$ & $19.96\pm0.02$ & $20.6^{+0.2}_{-0.2}$    &  $21.4^{+0.2}_{-0.3}$  \\            
4540  & $18.17\pm0.04$ & $17.96\pm0.02$ & $18.58\pm0.01$ & $18.80\pm0.01$ &  \phn$19.07^{+0.06}_{-0.07}$    &  $\phn19.29^{+0.05}_{-0.05}$  \\             
4581  & $18.47\pm0.04$ & $18.74\pm0.03$ & $19.82\pm0.02$ & $20.41\pm0.03$ & $21.6^{+0.3}_{-0.5}$    &  ...        \\
4589  & $19.77\pm0.09$ & $20.18\pm0.06$ & $21.34\pm0.05$ & $21.86\pm0.05$ & $\mathit{24.1^{+1.0}_{-1.0}}$    &  ...      \\
4970  & $19.63\pm0.08$ & $19.95\pm0.06$ & $20.96\pm0.04$ & $21.53\pm0.05$ & $22.4^{+0.4}_{-0.5}$    &  ...        \\
5020  & $18.09\pm0.04$ & $18.19\pm0.03$ & $19.21\pm0.02$ & $19.71\pm0.02$ & $20.7^{+0.2}_{-0.2}$    &  $\mathit{22.7^{+0.6}_{-1.6}}$    \\        
5325  & $19.78\pm0.09$ & $19.96\pm0.05$ & $20.98\pm0.04$ & $21.63\pm0.04$ & $22.5^{+0.4}_{-0.5}$    &  ...             \\
5350  & $17.63\pm0.03$ & $17.50\pm0.02$ & $18.49\pm0.01$ & $18.97\pm0.01$ & $19.9^{+0.1}_{-0.1}$    &  $\mathit{21.4^{+0.3}_{-0.4}}$    \\         
5379  & $19.87\pm0.09$ & $19.13\pm0.04$ & $19.23\pm0.02$ & $19.26\pm0.02$ & \phn$19.27^{+0.05}_{-0.06}$  &  \phn$19.31^{+0.04}_{-0.04}$     \\          
5434  & $19.00\pm0.06$ & $19.36\pm0.04$ & $20.53\pm0.03$ & $21.13\pm0.04$ & $\mathit{23.4^{+0.9}_{-0.9}}$    &  ...            \\
5671  & $18.80\pm0.05$ & $19.09\pm0.04$ & $20.17\pm0.02$ & $20.73\pm0.03$ & $22.0^{+0.4}_{-0.6}$    &  ...            \\
8104  & \phn$19.69\pm0.02^{\mathrm{c}}$ & $21.27\pm0.11$ & $22.15\pm0.07$ & $23.01\pm0.09$ & $23.1^{+0.4}_{-0.6}$    &  ...          \\ 
\hline
\multicolumn{7}{l}{$^{\mathrm{a}}$\textit{GALEX} NUV magnitude (AB system), except where noted \citep{galex}.}\\
\multicolumn{7}{l}{$^{\mathrm{b}}$Magnitudes in italic text are less than 3$\sigma$ detections.}\\
\multicolumn{7}{l}{$^{\mathrm{c}}$WFC3 F218W instrumental magnitude (this study).}\\
\end{tabular*}
\end{table*}

\subsection{WFC3 F218W Photometry}
Accurate modeling of the BSS population is a key factor in determining the amount of expected FUV emission.  We take special care  
to match modeled and observed near-UV (NUV) photometry (see Section~\ref{sec:bsmodel}). We use \textit{GALEX} NUV photometry 
for this purpose \citep{galex}. However, one BSS in NGC 188, WOCS 8104, is not detected in \textit{GALEX}. 
We obtained NUV photometry for WOCS 8104 using the \textit{HST} Wide 
Field Camera 3 (WFC3) in F218W. The observation occurred on 2012 June 19 as part of this program, with a total exposure 
time of 669~s. We use \textsc{daophot} for source detection and aperture photometry, and calculate 
the instrumental F218W magnitude for WOCS 8104 (given in Table~\ref{tab:bssphot}).

\section{ANALYSIS}
\label{sec:analysis}
\subsection{The FUV Color-Magnitude Diagram}
Identifying WD companions relies on detecting a FUV excess above the emission expected for a BSS alone.
We first compare the observed photometry with modeled BSS photometry for the entire BSS population. This 
approach best demonstrates the general trend of FUV emission for the type of BSSs found in NGC 188 in comparison to 
the FUV emission for potential BSS+WD pairs. It also clearly reveals those BSSs with highly significant FUV excesses 
indicative of hot (temperatures greater than 12,000 K) WD companions.

\begin{figure*}
\begin{center}
\includegraphics[scale=0.9]{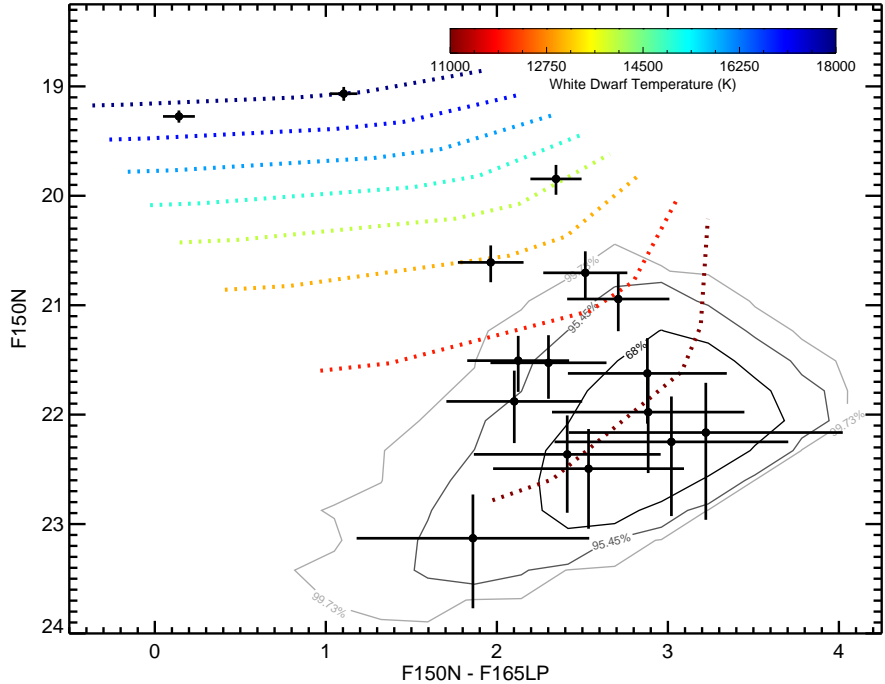}
\end{center}
\caption{FUV CMD of the NGC 188 BSSs, shown in black with 1$\sigma$ error bars. The three BSSs with less than a 3$\sigma$ detection in 
F150N are not included in this figure. The BSS--only model distribution is 
shown with grey contours at the 1, 2, and 3$\sigma$ level, as labeled in the figure. The rainbow-colored tracks show synthetic photometry 
of representative BSS+WD spectra with dotted lines showing tracks of constant WD temperature. The WD temperatures range 
from 11,000 to 18,000 K, as indicated by the color bar, with BSS temperatures varying from 5750 K (red end) to 6750 K (blue end) along each dashed line.
Four BSSs are clearly seen to have a significant FUV excess and are separated from the BSS--only distribution by more than 3$\sigma$ \citep[see also][]{Gosnell14}.}
\label{fig:uvcmd_all}
\end{figure*}

In Figure~\ref{fig:uvcmd_all} we plot a FUV color-magnitude diagram (CMD) for the BSSs in our study along with 
photometric models of the total BSS population and BSS+WD pairings (described in detail in the next two sections). 
The observed BSS photometry is shown with black points
and 1$\sigma$ error bars.
Since we use F150N--F165LP as our color diagnostic we exclude any BSS with less than a 3$\sigma$ detection 
in F150N in Figure~\ref{fig:uvcmd_all} and in all subsequent analyses.  This removes WOCS 4589 and 5434, 
both binary stars, and WOCS 1366, a non-velocity variable star. The lack of a F150N detection in these 
three sources is consistent within 3$\sigma$ for each of the BSSs 
given their faint NUV magnitudes.

The remaining BSS sample used for further analysis includes 16 sources: 13 SB1 binaries and three non-velocity variable stars.
The only three sources also having more than a 3$\sigma$ 
detection in F140N (WOCS 4540, 4348, and 5379) were presented in \citet{Gosnell14}, and are included in this sample of 16 BSSs.

\subsection{Modeling the Blue Straggler Star Population}
\label{sec:bsmodel}
We calculate the FUV emission of the overall NGC 188 BSS population 
using a population synthesis approach including a large number of model spectra spanning the entire range of 
observed BSS temperatures and luminosities. 
The modeling is done through a three step process: (1) we establish the ranges of physical parameters 
that define the NGC 188 BSS population, (2) we create a large sample of model spectra constrained 
by those ranges, and  
(3) we calculate synthetic photometry for the sample of model spectra. 

First, we define the range of parameters for the model spectra using the known 
temperatures and luminosities of the NGC 188 BSSs. 

The BSS temperatures are determined using metallicity-dependent 4- or 5-band color-to-$T_{\mathrm{eff}}$ conversions 
from \citet{Ramirez05}. The colors used are $B-V$, $V-J$, $V-H$, and $V-K$ \citep{Sarajedini99,2mass}.  $V-I$ is also used 
when $I$ photometry is available. The resulting temperatures and errors for the non-velocity variable and SB1 BSSs  
are given in Table~\ref{tab:bssinfo}.  The errors given include the error in each color measurement and the systematic error in the 
temperature calculation as given in \citet{Ramirez05}.

We set the luminosity range based on 
\textit{GALEX} NUV magnitudes for the BSSs, which vary from 20.6 $M_{\mathrm{NUV,AB}}$ to 
18.0 $M_{\mathrm{NUV,AB}}$ \citep{galex}. As \textit{GALEX} NUV photometry is the bluest luminosity 
information available it provides the most accurate normalization for estimating the FUV flux. 

Next, we create 50,000 individual model spectra that, together, constitute the BSS--only model population. 
We create smooth probability density functions (PDFs) modeled after the observed 
shapes of BSS temperature and NUV luminosity 
cumulative distribution functions (CDFs). 
We Monte Carlo sample the PDFs to determine the parameters used 
for each model spectrum.
There is only a weak correlation between 
temperature and luminosity in the NGC 188 BSS population ($r=-0.2$), so the two parameters are sampled 
independently. For each sampling of the PDFs we calculate a model spectrum by interpolating between reddened 
UVBLUE spectra 
with temperatures of 6000, 6500, 6750, and 7000 K to match the sampled temperature 
\citep[E($B-V$) = 0.09, {[Fe/H]} = 0, $\log (g)=3.5$;][]{Sarajedini99,uvblue}. We then normalize the interpolated spectrum 
to match the sampled NUV luminosity. The interpolation and normalization steps are carried out  
on all 50,000 PDF samples. As a result, we have a BSS--only model composed of a large set 
of model spectra spanning the range of physical parameters observed in the NGC 188 population. 

Finally, after creating the BSS--only model we calculate the expected FUV CMD for the entire  
model distribution. We convolve each model spectrum with \textsc{synphot} F140LP, F150LP, and F165LP 
throughput curves to obtain count rates. 
Since the observed data are red leak-corrected the throughput curves do not include a red leak component. We add 
Poisson noise to the convolved count rates to mimic photometric errors. The count rates and flux errors in F140N and F150N are 
calculated in the same manner as the observations. We calculate magnitudes in F140LP, F150LP, F165LP, F140N, and F150N 
for each of the 50,000 modeled spectra. 

We compare the modeled BSS--only FUV photometry with the observed BSSs in Figure~\ref{fig:uvcmd_all}. The 2-dimensional density 
histogram of modeled photometry is represented by the gray contours encompassing 68\%, 95.45\%, and 99.73\% of the 
distribution, as labeled. We restrict the density 
contours to those modeled sources that meet or exceed the same 3$\sigma$ F150N detection threshold we apply to the observations
\footnote{We note that a similar figure in \citet{Gosnell14} does not have the same 3$\sigma$ detection cutoff, resulting in 
contours that extend fainter and redder than the contours shown here. In either case, the extent of the contours into the lower right 
corner of the CMD does not impact our search for WD companions.}.

\subsection{Sources with Significant FUV Excesses}
\label{sec:fuvsources}

In order to demonstrate how the presence of a WD companion changes the expected FUV emission we create 
representative pairs of BSS+WD binaries. Keeping the BSS luminosity constant at a \textit{GALEX} NUV magnitude 
of 18.0 to mimic the brightest BSS in the NUV, 
we add together the spectra of BSSs between 5750 K and 6750 K \citep{uvblue} with WDs of increasing temperature, ranging 
from 11,000 K to 18,000 K. We use WD spectra with a $\mathrm{log} (g) = 7.75$ \citep[P. Bergeron, private communication;][]{Wood95}. 
We calculate the 
synthetic photometry for each BSS+WD 
pair using the same method as for the BSS--only distribution.

The synthetic photometric results for the BSS+WD pairs are shown in Figure~\ref{fig:uvcmd_all} as dotted lines. The 
rainbow colors indicate the WD temperature, as shown with the color bar. The length of each line shows the photometry of 
that particular WD temperature with a BSS of 5750 K on the left (blue end) extending to a BSS of 6750 K 
on the right (red end).  Since a higher 
temperature BSS contributes more light in F165LP it results in an overall redder color. The position of the rainbow tracks 
changes slightly with the choice of BSS luminosity, but a single BSS luminosity is sufficient to illustrate the 
general photometric trends for BSS+WD binaries.

Visual analysis of Figure~\ref{fig:uvcmd_all} shows there are four sources with obvious FUV excesses: WOCS 4348, WOCS 4540, 
WOCS 5350, and WOCS 5379. These sources are well separated from the BSS--only distribution and their photometry is consistent 
with the FUV excesses expected for BSS+WD binaries. 

\begin{figure*}
\begin{center}
\includegraphics[scale=1.0]{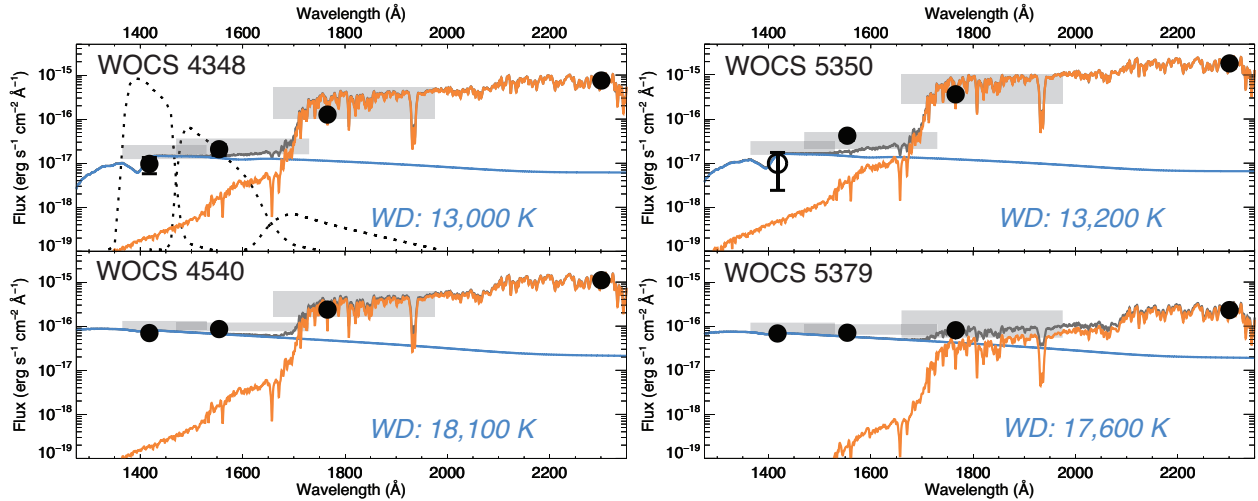}
\end{center}
\caption{Best-fit BSS+WD spectra for the four BSS with hot WD companions, as labeled.
The WD spectrum is shown in blue with the temperature
as labeled, the mean BSS temperature spectrum (given in Table~\ref{tab:bssinfo}) is 
shown in orange, and the combined BSS+WD spectrum is shown 
in dark gray. The synthetic photometry of the combined spectrum is shown 
with the light gray boxes. The vertical extent of the boxes shows the photometry range given 
the temperature uncertainty of the BSS and WD. The horizontal extent of the boxes represents the filter width,
including the red tail of each bandpass. 
The upper left plot also includes the effective narrow-band throughputs (black dashed lines) as a guide. 
The observed photometry in F140N, F150N, F165LP, and \textit{GALEX} 
NUV is shown as filled black circles at the reference wavelength for each bandpass. 
Measurements in F140N that are below 3$\sigma$ are shown 
as an open circle.  Observed flux error bars (3$\sigma$) are plotted in black when the error exceeds the size of the symbol. 
The best-fit spectrum is required to match the \textit{GALEX} NUV measurement. 
}
\label{fig:specphot}
\end{figure*}

We fit BSS+WD spectra to each of the four FUV excess sources individually to estimate the companion WD temperature. 
In Figure~\ref{fig:specphot} we show the F140N, F150N, F165LP, and \textit{GALEX} NUV photometric 
data for these sources along with 
the best-fit BSS+WD spectrum found through weighted least squares minimization. 
A similar figure for WOCS 4348, 4540, and 5379 is shown in \citet{Gosnell14}, although here we 
employ an improved method of finding the best-fit WD temperature. The grid of WD models (P. Bergeron,
private communication)  
has a temperature resolution of 1,000 K.  In \citet{Gosnell14} we restricted our search to 
grid temperature values while in this work we interpolate between temperatures. Given the uncertainty 
in the best-fit WD temperatures, the results here and in \citet{Gosnell14} are consistent.
In each case the mean temperature BSS spectrum is shown in orange, the WD spectrum in blue, and the combined spectrum in gray. 
The synthetic photometry of the combined spectrum is shown with the light gray boxes. The horizontal extent of the boxes
shows the filter width while the vertical extent shows the range in photometry, 
including the BSS and WD temperature uncertainties.  
The combined spectrum is required to fit the \textit{GALEX} NUV photometry. 
The narrow-band effective throughputs are shown in the upper left plot as a guide.  
The observed photometry is shown in black circles at the reference wavelength for each bandpass, 
and observed 3$\sigma$ photometric errors are shown when 
they exceed the symbol size. Measurements in F140N that are less than 3$\sigma$ are shown with an open circle. 
The best-fit spectra are also checked for consistency with $U$-band photometry, when available \citep{Sarajedini99}.
 
We use the estimated temperature and uncertainty from Figure~\ref{fig:specphot} to find the range 
of WD ages for each source \citep{Holberg06,Tremblay11}. In Table~\ref{tab:wdtemps} we list the 
best-fit WD temperature estimates and the age range for each of the four BSSs with a hot WD companion, along with 
the observed binary period and eccentricity \citep{Geller09}. Of these BSSs, the maximum age is $230\pm30$ 
Myr for WOCS 4348, so all four BSSs 
formed very recently in comparison to the 7 Gyr age of NGC 188. 

 \begin{table*}[]
\begin{center}
 \caption{Hot WD companion temperature and age estimates} \label{tab:wdtemps} 
 \vspace{3 mm}
 \begin{tabular*}{400pt}{c @{\extracolsep{\fill}} cccc}
 WOCS ID  & WD temp (K) & Age (Myr)$^{\mathrm{a}}$ & Period (days)$^{\mathrm{b}}$ & eccentricity$^{\mathrm{b}}$ \\ 
 \hline
 4348 & $13000\pm500$ & $230\pm30$ & $1168\phantom{.00}\pm\phantom{0}8\phantom{.00}$  & $0.09\pm0.05$ \\
 4540 & $18100\pm500$ & $70\phantom{0}\pm\phantom{0}7$ & $3030\phantom{.00}\pm70\phantom{.00}$ & $0.36\pm0.07$ \\
 5350 & $13200\pm500$ & $220\pm30$ & $\phantom{0}690\phantom{.00}\pm\phantom{0}3\phantom{.00}$ & $0.07\pm0.05$ \\
 5379 & $17600\pm500$ & $77\phantom{0}\pm\phantom{0}7$ & $\phantom{0}120.21\pm\phantom{0}0.04$  & $0.24\pm0.03$ \\
 \hline
 \multicolumn{4}{l}{$^{\mathrm{a}}$\citet{Holberg06,Tremblay11}.}\\
 \multicolumn{4}{l}{$^{\mathrm{b}}$\citet{Geller09}.}\\
 \end{tabular*}
\end{center}
 \end{table*}

\subsection{The FUV-Faint BSS Sample}

There are 12 remaining BSS (nine binaries, three non-velocity variables) within the sample of F150N sources  
that lack a significant (greater than $3\sigma$) FUV excess. The possibility remains that some 
of these BSSs have a cool WD companion between 11,000--12,000 K.  
At these temperatures the WD emission would shift the observed photometry to be 
slightly bluer and brighter than the expected emission for that particular BSS.  We 
cannot detect any WDs with temperatures below 11,000 K. Such ``cold'' WDs would not have enough FUV 
emission on their own to exceed the expected emission for any BSS in NGC 188.

We investigate whether any of the 12 FUV-faint BSSs contain a cool (11,000--12,000 K) WD 
companion by modeling the expected photometry for each BSS individually.  Following  
a similar methodology as in Section~\ref{sec:bsmodel}, for each individual FUV-faint BSS we do a 
Monte Carlo sampling of the 
temperature range given in Table~\ref{tab:bssinfo} 
and luminosity range, as constrained by \textit{GALEX} NUV photometry for that BSS.  The 
resulting sample is synthetically observed to create a 2-dimensional density distribution  
of expected FUV emission in F150N and F150N$-$F165LP for that particular BSS.  
We then find the model density value with the same color and magnitude as the BSS 
observation, and adopt that value as the probability the BSS is consistent with the expected FUV 
emission of a BSS without a WD companion ($P_{\mathrm{BSS}}$).  
The probability that the BSS has a WD companion is then $P_{\mathrm{WD}} = 1 - P_{\mathrm{BSS}}$.

The modeled photometry for each BSS is highly dependent on BSS temperature, which is more uncertain than 
the NUV luminosity used to normalize the model spectra.  We compared the 
temperatures used here with recently calculated temperatures found from H$\alpha$ line-profile 
fits (K. Milliman, private communication).  Many BSS temperatures are consistent across the methods, but 
WOCS 4230 and 4581 have temperature differences of approximately 200 K.  Additionally, the temperatures of 
the non-velocity variable BSSs WOCS 4306 and 5020 vary by 200--300 K.  If the actual BSS temperature is different 
from the modeled BSS temperature the observed BSS photometry could mimic the presence of a cool WD companion.  
In order to account for potential offsets in the known temperatures, and avoid false identification of a WD companion, 
we also model each BSS with 
temperature ranges shifted by $\pm200$ K and adopt the highest $P_{\mathrm{BSS}}$, or lowest $P_{\mathrm{WD}}$, 
for each BSS.

\begin{table*}[]
\footnotesize
\begin{center}
 \caption{FUV-faint BSSs: Probability of a cool WD companion} \label{tab:wdprob} 
 \vspace{3 mm}
 \begin{tabular*}{450pt}{r @{\extracolsep{\fill}} cccccccccccc}
 \hline
 WOCS ID :  & 451 &  1888  & 2679  & 4230  & 4290 & 4306 & 4581 & 4970 &  5020 & 5325 & 5671 & 8104  \\ 
 $P_{\mathrm{WD}}$: & 0.13 & 0.96  & 0.97  &  0.99 & 0.63 & 0.74 & 0.56 & 0.79 & 0.92 & 0.81 & 0.70 & 0.10  \\    
 \hline
 \end{tabular*}
\end{center}
 \end{table*}

\begin{figure*}
\begin{center}
\includegraphics[scale=1.0]{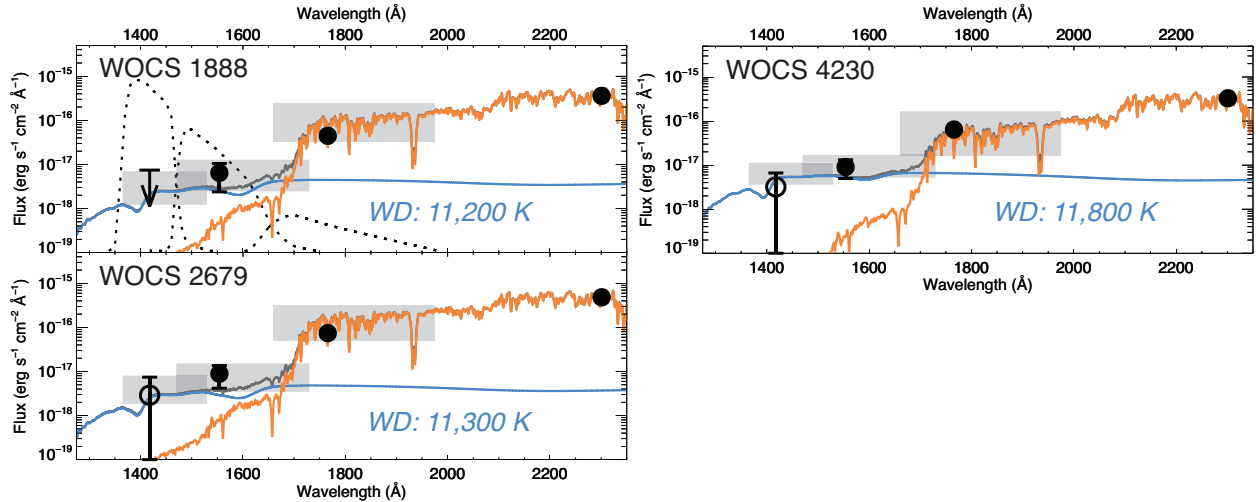}
\end{center}
\caption{Best-fit BSS+WD spectra for the three BSS with cool (11,000--12,000 K) WD companions, as labeled.  The colors and symbols used are 
the same as in Figure~\ref{fig:specphot}. For sources without a F140N detection the 3$\sigma$ upper limit is shown 
with a black bar and a down arrow. Due to the fainter WD flux these three BSSs have 
WD companions detected 
at the 2$\sigma$ level.  The WD temperatures here correspond to ages between 310--360 Myr \citep{Holberg06,Tremblay11}.}
\label{fig:2sigWDs}
\end{figure*}

In Table~\ref{tab:wdprob} we list each FUV-faint  
BSS along with the probability that the source has a cool WD companion 
between 11,000--12,000 K. 
There are three BSS with probabilities greater than $2\sigma$: 
WOCS 1888 ($P_{\mathrm{WD}} = 96$\%), 
WOCS 2679 ($P_{\mathrm{WD}}= 97$\%), and WOCS 4230 ($P_{\mathrm{WD}} = 99$\%).  
The photometry for WOCS 1888, 2679, and 4230 are best fit with BSS+WD composite spectra with 
WD temperatures between 11--12,000 K, as shown in Figure~\ref{fig:2sigWDs}. The 
best-fit temperatures and corresponding 
ages for these sources, along with binary orbit parameters, are given in Table~\ref{tab:coolwdtemps}.
There is a 2\% chance 
that one of the 12 FUV-faint BSSs has a probability equal to or greater than 
96\% due to random fluctuations alone, but only a 0.03\% chance that all three are due to 
random fluctuations. We assume that these three BSSs have WD companions and 
formed through mass transfer between 310--360 Myr ago \citep{Holberg06,Tremblay11}.

 \begin{table*}[]
\begin{center}
 \caption{Cool WD companion temperature and age estimates} \label{tab:coolwdtemps} 
 \vspace{3 mm}
 \begin{tabular*}{450pt}{c @{\extracolsep{\fill}} cccc}
 WOCS ID  & WD temp (K) & Age (Myr)$^{\mathrm{a}}$ & Period (days)$^{\mathrm{b}}$ & eccentricity$^{\mathrm{b}}$ \\ 
 \hline
 1888 & $11200\pm500$ & $360\pm50$ & $2240\pm30$ & $0.21\pm0.04$   \\
 2679 & $11300\pm500$ & $350\pm50$ & $1033\pm\phantom{0}8$ & $0.07\pm0.05$ \\
 4230$^{\mathrm{c}}$ & $11800\pm500$ & $310\pm40$ & ... & ... \\
 \hline
 \multicolumn{4}{l}{$^{\mathrm{a}}$\citet{Holberg06,Tremblay11}.}\\
 \multicolumn{4}{l}{$^{\mathrm{b}}$\citet{Geller09}.}\\
 \multicolumn{4}{l}{$^{\mathrm{c}}$WOCS 4230 is a binary, but does not have a solved binary orbit }\\
 \multicolumn{4}{l}{due to rapid rotation \citep{Geller09}.}\\
 \end{tabular*}
\end{center}
 \end{table*}

\begin{figure*}
\begin{center}
\includegraphics[scale=1.0]{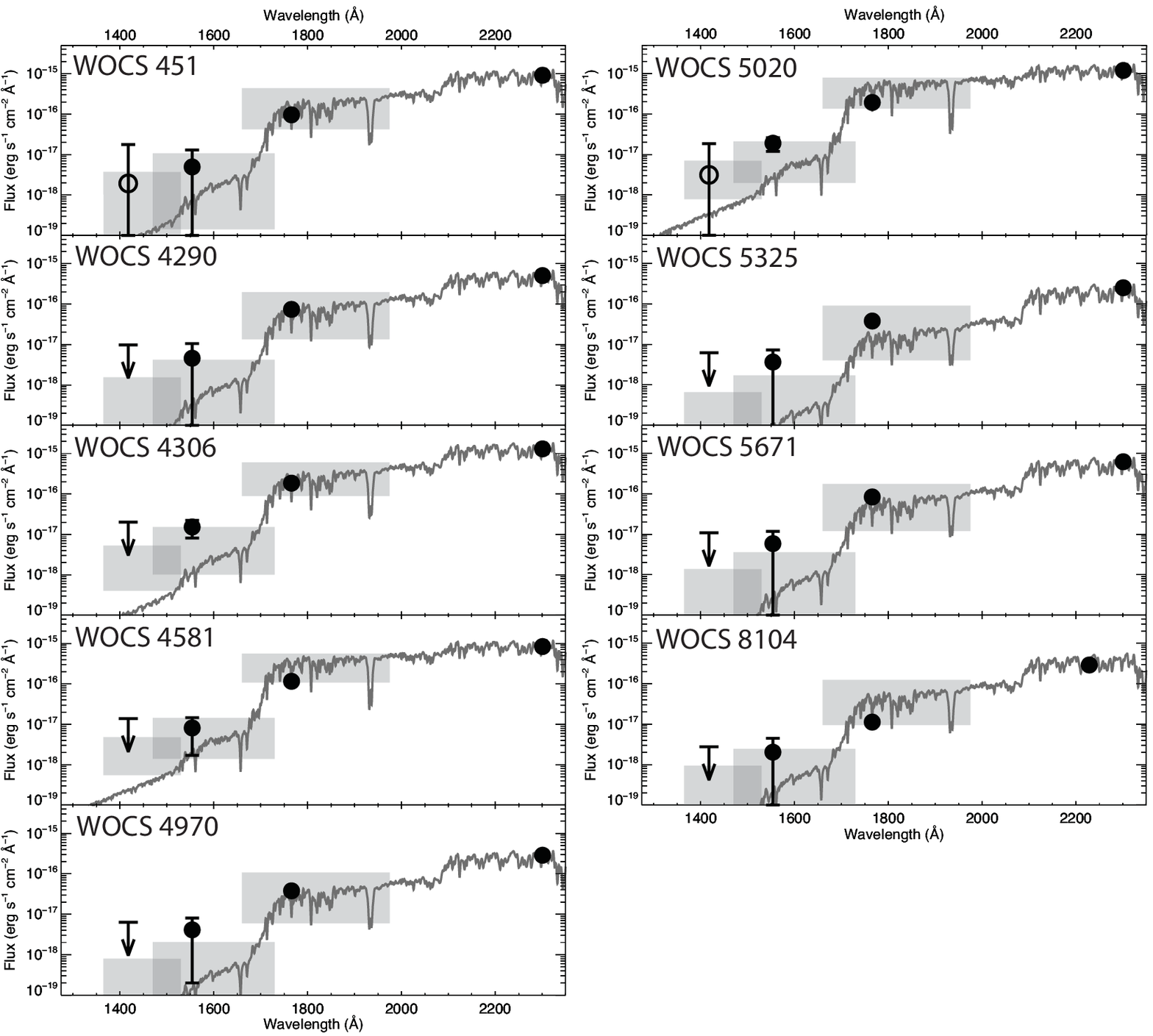}
\end{center}
\caption{Best-fit spectra for the BSSs without significant FUV excesses.  The colors and symbols used are 
the same as in Figures~\ref{fig:specphot} and~\ref{fig:2sigWDs}. These sources do not have detected 
WD companions with temperatures of 11,000 K
or greater. (The NUV point for WOCS 8104 is WFC3 F218W photometry from this study.)}
\label{fig:noWDs}
\end{figure*}

For the remaining FUV-faint BSSs, the fit between the observed and modeled photometry is not improved by adding 
a cool WD of any temperature greater than 11,000 K.  The photometry of these sources is consistent with a BSS-only synthetic spectrum, 
and these sources do not have WD detections. The best fit spectra for the nine remaining BSSs with the 
corresponding \textit{HST} and \textit{GALEX} photometric points are shown in Figure~\ref{fig:noWDs}. 

\section{THE NGC 188 BSS POPULATION}
\label{sec:bspop}

The \textit{HST} photometry reveals four hot WD companions hotter than 12,000 K and three 
additional WD companions between 11,000--12,000 K among 15 BSS SB1 binaries 
and four non-velocity variable BSSs. All of the detected WDs are in 
known binary systems.
These data do not 
preclude the presence of cold WD companions (temperatures less than 11,000 K) 
among the remaining BSS, but we are unable to identify 
cold WD companions using these photometric data. 

\begin{figure}[!h]
\begin{center}
\includegraphics[scale=0.75]{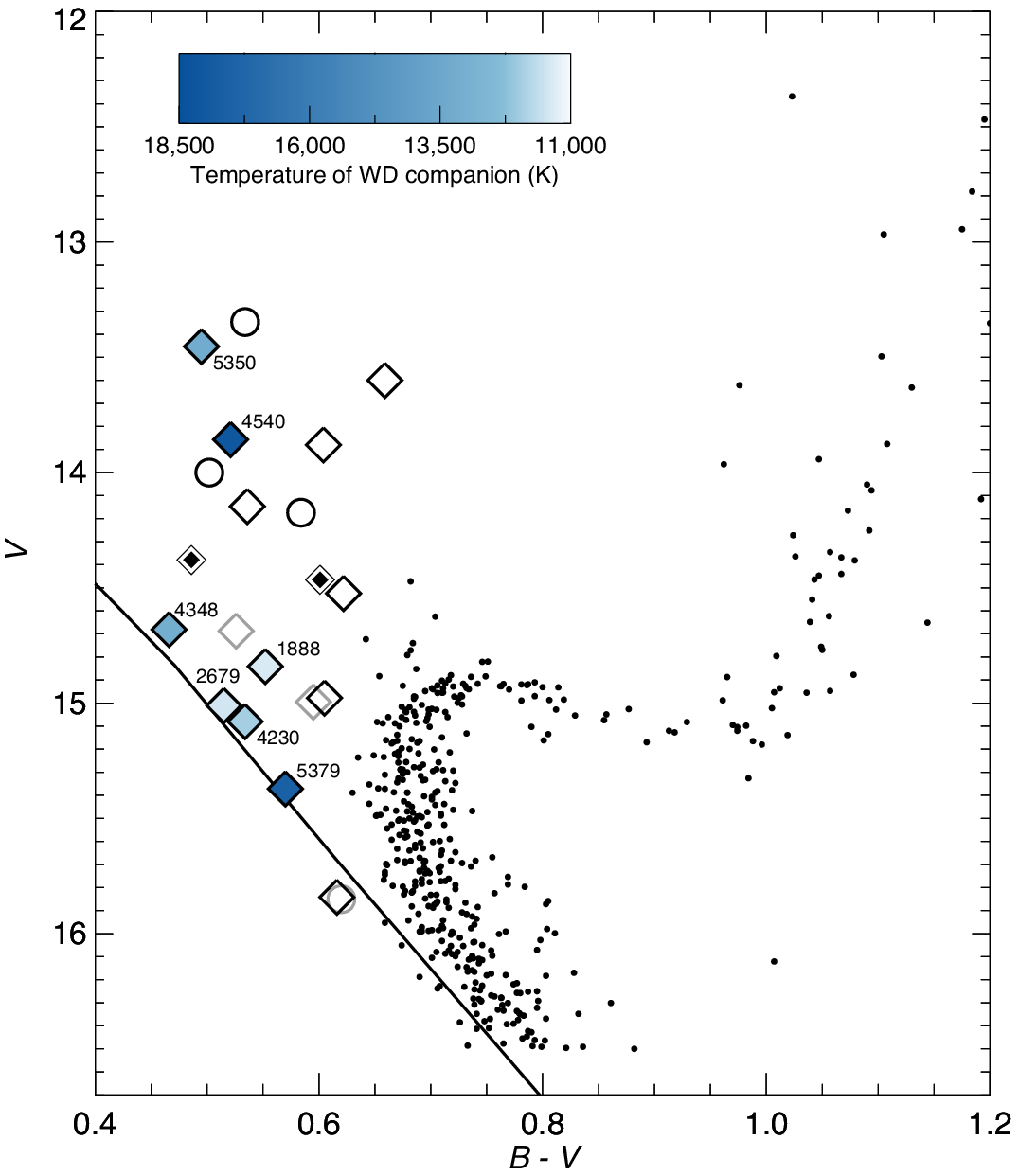}
\end{center}
\caption{Optical CMD 
of NGC 188 cluster members with the BSS population highlighted according to binarity and the measured temperature of
WD companions. The solid black line is the ZAMS for NGC 188. Binary BSSs are shown as diamonds and 
non-velocity variable BSSs are shown as large circles. The outlined solid black diamonds are the two 
double-lined BSS binaries that are not included in this 
\textit{HST} study. The BSSs with WD detections are shown with 
a color from dark blue to light blue representing the temperature of the WD companion, as indicated with the color bar.  
The sources outlined in grey (two SB1 binaries, one single BSS) are 
the three sources without a 3$\sigma$ detection in F150N. }
\label{fig:cmdbs}
\end{figure}

In Figure~\ref{fig:cmdbs} we plot an optical CMD ($V$, $B-V$) of NGC 188 member stars \citep{Geller08}. The cluster 
zero-age main sequence (ZAMS) is shown as a solid black line. The BSSs 
are highlighted according to their binarity and WD companion temperature, which is a proxy for BSS age. Binary BSSs are plotted as 
diamonds and non-velocity variable BSSs are plotted as large circles. The two SB2 BSSs, not part of this \textit{HST} 
study, are shown as outlined black diamonds. The BSS binaries with WD detections are shown with a color from dark blue 
to light blue representing the measured WD temperature, as indicated with the color bar.
The three BSSs with less than 
a 3$\sigma$ detection in F150N are shown outlined in light grey instead of black. 

Since WD temperature correlates with age, we can use the CMD location of BSSs with detected WD companions to 
probe relationships between BSS age 
and optical luminosity and color. Several of the youngest BSSs (dark blue diamonds) sit  
along the ZAMS. BSS binaries along the ZAMS include two of the BSSs with hot WD companions (WOCS 4348 and 5379),  
two of the BSSs with cool WD companions 
(WOCS 2679 and 4230), and an additional binary with a 10\% 
probability of having a WD companion (WOCS 8104).  
Based on the detected WD temperatures, the spread of ages along the ZAMS 
is 77--360 Myr \citep{Holberg06,Tremblay11}, and may extend further if 
WOCS 8104 has a cold and old WD companion.

At the same time, two of the youngest mass transfer-formed BSSs (WOCS 4540 and 5350) 
lie well off the ZAMS. 

These results suggest that BSS 
products along the ZAMS may in fact be among the youngest in the population, \textit{but distance 
from the ZAMS is not necessarily equivalent with age since formation}. One must use caution 
when inferring the age of BSSs using single-star isochrones, especially for BSSs that lie away from the ZAMS. 

Three of the four non-velocity variable BSSs are also among the 
most optically luminous of the population. If these BSSs formed through collisions they would likely 
be among the most massive BSSs, 
perhaps accounting for their high optical luminosity \citep{Leigh11}.  And yet, they are not more luminous 
than the most luminous mass transfer-formed BSSs. There is not a clear color or luminosity separation between mass transfer 
BSSs and possible collision BSSs in the optical CMD.  

\subsection{Frequency of the Mass Transfer Formation Channel}
\label{sec:MTfrequency}
Previous work predicted that the majority of the NGC 188 binary BSSs 
formed through mass transfer \citep{Mathieu09, Geller11}.  
The seven WDs detected in this work indicate that more than half 
of the SB1 BSSs formed through mass transfer. Older mass
transfer-formed BSSs have cold WD companions that cannot be detected 
with this photometric method, so the total number of 
mass transfer-formed BSSs is greater than seven. Therefore, our results confirm previous 
predictions that the majority of binary BSSs in NGC 188 formed through mass transfer.

We can further constrain the total frequency of mass-transfer formation by 
considering how many BSSs may form through other formation mechanisms.

We do not expect any of the non-velocity variable BSSs to form through Case B or C Roche lobe overflow 
mass transfer. As non-velocity variable objects, these BSSs are either single stars or 
binary stars with extremely long periods greater than 10,000 days.
Case C mass transfer systems have final orbital periods less than about 3,000 
days \citep{Chen08}, which is well within the RV binary detection limit in NGC 188 \citep{Geller09}. 
(We note that there is a small possibility for a non-velocity variable source to be in a face-on 
binary that is undetectable using radial velocity methods. Incompleteness studies indicate there 
is a 24\% chance that one of the non-velocity variable BSSs in NGC 188 is in a binary with 
a period less than $10^{4}$ days but is not detected due to inclination; \citealt{Geller12}.)
We propose that 
the non-velocity variable BSSs formed through other mechanisms, such as collisions 
or mergers from contact binaries. 

Is it possible for any of the remaining eight SB1 BSS binaries to form through non-mass transfer processes? 
It is unlikely that the SB1 BSSs formed through collisions. BSSs formed through collisions can be 
in binaries, but the orbital periods tend to be very long and beyond RV sensitivity limits \citep{Geller13}. 
In addition, given the observed statistical secondary mass distribution the SB1 BSS secondary masses 
are between 0.2--0.8 $M_{\odot}$ \citep{Geller11}, with a strong peak at 0.5 $M_{\odot}$. 
Models of collisionally-formed BSS 
binaries result in more massive companions than the observed 
secondary mass distribution allows \citep{Geller11}.

Some SB1 BSSs may have formed through Kozai-induced mergers 
in a hierarchical triple \citep{Perets09,Naoz14}. The resulting period distribution would be similar 
to that observed in the NGC 188 BSS binaries, but the secondary masses would differ \citep{Perets14}. 
The original tertiary, now the BSS secondary, would be drawn from the cluster initial-mass 
function \citep{Kroupa01}. 

Without strong observational constraints on the primordial triple population in clusters it is 
difficult to make confident predictions about the number of Kozai-formed BSSs. Nonetheless, 
\citet{Geller13} put limits on Kozai-BSS formation with an $N$-body model of NGC 188 containing 
200 primordial triples --- essentially 
giving every primordial binary with periods between 2 and 50 days a tertiary companion. This model 
creates an additional 1--2 BSSs at the age of NGC 188, but most of the BSSs result from collisions involving  
triple systems, not from the Kozai mechanism itself. Approximately 0.5 BSSs per model are created 
through the Kozai mechanism.  However, the \citet{Geller13} Kozai model 
prescriptions are possibly inaccurate such that the Kozai origin could be more prevalent than 
current $N$-body models suggest.
We estimate that one of the BSS SB1 binaries may have formed through a merger 
in a hierarchical triple system. 

We currently have no specific hypothesis for the formation mechanism of the two
SB2 short-period BSS binaries. Case B and C mass transfer cannot result in a short-period 
BSS binary with a mass ratio near unity, as seen in the case of WOCS 5078 and 7782. $N$-body models 
fail to create similar systems, even with populations of collisionally formed BSSs 
\citep{Hurley05,Geller13}. These BSSs may have formed through any of the potential formation 
pathways and exchanged into their current binaries, but there is no way to know for certain. 
As they have more complex origins we exclude them from the following discussion 
of broad BSS population characteristics.

In summary, we detect WD companions to 7 of the 15 SB1 BSSs, establishing a lower limit of 
the mass-transfer formation frequency of 33\%.  
Given the binary properties and inferred companion masses of the remaining SB1 BSSs in NGC 188, 
our investigation of the currently hypothesized formation mechanisms suggests that 14 SB1 BSSs were formed through mass transfer.  
Thus, 14 of the 21 BSSs likely formed through mass transfer, for a 
total BSS mass-transfer formation frequency in NGC 188 of approximately 67\%.

\subsection{Consistency with Binary Population Synthesis Models}

Now that other BSS formation scenarios have been considered we can make an  
appropriate comparison between our empirical results and the theoretical 
results of the \citet{Geller13} $N$-body model of NGC 188.  
\citet{Geller13} present 20 full-lifetime $N$-body realizations of NGC 188, with complete tracking of the 
binary populations and dynamical encounters. The models create mass transfer-formed BSSs, and 
retain information on when the mass transfer ended in each case. 

\begin{figure}
\begin{center}
\includegraphics[scale=0.5]{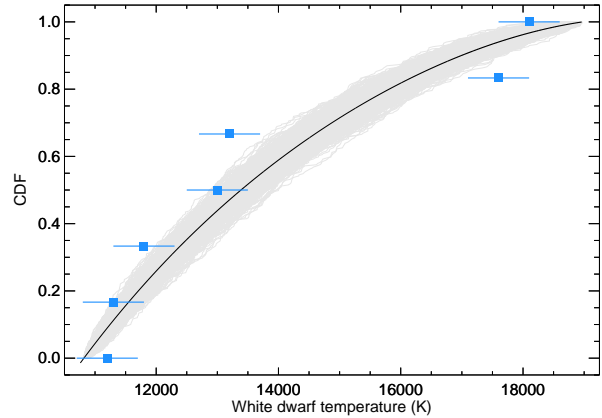}
\end{center}
\caption{CDF of detected WD 
temperatures compared to $N$-body model predictions. The distribution from 
the $N$-body model of NGC 188 is shown as a black line \citep{Geller13}. The light gray lines are from 1000 Monte Carlo 
bootstrap resamples of the $N$-body distribution.  The seven WDs detected in this study are shown as solid blue squares at the 
temperatures given in Figures~\ref{fig:specphot} and~\ref{fig:2sigWDs} with 500 K error bars. A KS test indicates the CDF 
of the detected WD temperatures is consistent with being drawn from the same parent population as the 
$N$-body model distribution ($p=0.97$).
}
\label{fig:tempcdf}
\end{figure}

With seven WD detections we can directly compare the empirical 
and theoretical age distributions for mass transfer-formed BSSs.
We create the $N$-body age distribution using all mass transfer-formed 
BSSs observable between 6.0--7.5 Gyr \citep{Geller13}, and convert the time since mass 
transfer ended to a WD temperature \citep{Holberg06,Tremblay11}.  The 
resulting CDF is shown in Figure~\ref{fig:tempcdf} as a solid black line.
The light gray lines are the result of Monte Carlo bootstrap resampling of the 
$N$-body CDF, meant to illustrate the inherent error in the $N$-body age 
distribution.

We compare the $N$-body age distribution with the empirical age 
distribution, shown in Figure~\ref{fig:tempcdf} as blue squares with 
500 K error bars.  A KS test shows the empirical distribution and
theoretical distribution are consistent with being drawn from the same 
parent distribution ($p=0.97$).  The overall shape of the $N$-body 
age distribution is fully consistent with the BSS ages detected in this study.

Adopting a mass-transfer formation frequency of 67\% for the BSS 
population implies there are five BSSs in NGC 188 that formed through other formation 
mechanisms, such as collisions or the 
Kozai mechanism, in addition to the two short-period SB2 systems. 
This \textit{number} is consistent with the number of collisionally formed BSSs 
seen in the \citet{Geller13} $N$-body model of NGC 188. 

The \textit{fraction} of collisionally formed BSSs in the \citet{Geller13} study is quite different 
than observed, however, because the total number of BSSs created is too low. The model 
results have an average of six BSSs in NGC 188 at 7 Gyr, one of which formed from mass transfer. 
A separate $N$-body study of the open cluster M67 also fails to create a high fraction of mass transfer-formed 
BSSs as observed in NGC 188 \citep{Hurley05}, although the progenitor binary period distributions used were not realistic.

The lack of BSSs in $N$-body models \citep{Geller13} may be attributed 
to an incomplete description of mass transfer processes in the binary population synthesis 
models used within the $N$-body code. \citet{Geller13} utilize the 
NBODY6 code \citep{nbody6}, which relies similar algorithms to those 
implemented in the Binary-Stellar Evolution (BSE) code of \citet{Hurley02} 
to track binary evolution. 

Although there are too few BSSs, \citet{Geller13} note that the model
produces a large number of long-period post-common envelope (CE) binaries that 
are not observed in such frequency in NGC 188 or the field. 
The mass transfer parameterization in BSE may need to be adjusted such that 
these sources go through stable Roche lobe overflow rather than CE.  
Converting the post-CE binaries to BSS+WD binaries 
would bring the total 
mass transfer BSS population to 10 systems at 7 Gyr, for a 
mass-transfer formation frequency of 67\%. This matches the inferred mass-transfer
formation frequency of 67\% measured in this study for the total NGC 188 BSS population.

However, 
if the parameterization of mass transfer is amended in BSE the 
consistency of the  
age distributions from the $N$-body model and 
and these results should be revisited.

Binary population synthesis models are important tools for theoretical 
astronomy that allow full-$N$ models of rich open clusters to run in a reasonable amount of 
time. We hope that observations such as these help constrain the parameterizations used 
so that population synthesis models can be as accurate as possible.

\section{SUMMARY}
\label{sec:summary}

We utilize a FUV photometric study of the BSS binaries in the old open cluster NGC 188 
to search for FUV excesses indicative of WD companions and thereby BSSs formed 
through mass transfer. We detect four BSSs with hot (temperatures greater than 12,000 K) 
WD companions. Since WDs cool as they age we use the estimated temperatures of the 
WD companions to place limits on the WD ages. These four systems are younger than 
250 Myr, indicating the mass transfer in these binaries ended very recently. By further 
comparing the expected and observed FUV emission on a star-by-star basis with a 
modeled Monte Carlo distribution of expected BSS emission, we detect three 
additional cool WD companions with temperatures between 11,000--12,000 K.

In total, the \textit{HST} data point to seven BSSs forming through mass transfer in the past 400 Myr. 
Analyzing the location of these recently-formed BSSs on 
an optical CMD reveals that these very young BSSs are both on and off the ZAMS. 
Thus, while BSSs near the ZAMS are likely among the youngest of the population, distance from the ZAMS 
is not a direct indicator of BSS age. Single-star isochrones may not be appropriate for determining 
the age of luminous BSSs. 
We do not find a clear separation in the optical CMD 
between mass transfer-formed BSSs and the non-velocity variable BSSs possibly formed through collisions
or mergers of contact binaries.

The seven detected WDs in this study set a lower limit of 33\% for the mass-transfer 
formation frequency of BSSs in NGC 188. 
In addition, we explore 
the possibility of other formation mechanisms by comparing overall properties of the BSS population 
to theoretical predictions.  Due to the BSS binary properties it is very unlikely 
that any of the SB1 BSSs formed through dynamical collisions. Collisions typically create either single BSSs 
or BSSs with long periods that are not detectable with RVs \citep{Geller13}. It is unlikely that any 
of the non-velocity variable BSSs formed through Case B or Case C mass transfer, as those mass transfer 
products are most often binaries with periods well within our RV sensitivity \citep{Chen08,Geller09}. 
Current $N$-body models \citep{Geller13} predict 
that the Kozai mechanism produces 0.5 BSSs at the age of NGC 188, so we 
assume that one SB1 binary may have formed via merger in a hierarchical triple 
via the Kozai mechanism.  As a result, we find that 
14 of the 15 SB1 BSSs likely 
formed through mass transfer.  This results in a total mass-transfer formation frequency of 
approximately 67\% (14 of 21) for BSSs in NGC 188. 
These results strongly support the previous predictions in \citet{Mathieu09} and \citet{Geller11} that 
the majority of the NGC 188 BSSs form through mass transfer.  

The distribution of BSS ages found in this study is fully consistent with the mass transfer-formed 
BSS age distribution of mass-transfer formed BSSs in the $N$-body simulations of \citet{Geller13}.  
However, the frequency of mass-transfer formed BSSs 
found here challenges the results of $N$-body models that have difficulty 
creating numerous mass transfer BSSs \citep{Hurley05,Geller13}. It is possible that the parameterization 
of unstable mass transfer causes too many BSS-progenitor binaries to go into a stage of CE rather than 
stable Roche lobe overflow. If the parameterization within the $N$-body model of NGC 188 
was adjusted such that the long-period CE systems instead went through stable mass transfer, the model 
would reproduce the BSS mass-transfer formation frequency measured in this study. 

These BSS+WD binary detections represent some of the most well-characterized 
post-mass transfer systems to date, and provide an unparalleled opportunity to 
constrain detailed mass transfer and binary evolution models \citep{Gosnell14}.  These BSS+WD 
binaries will also help address the discrepancy between observations and $N$-body and population 
synthesis results. 

\acknowledgements We are grateful to the anonymous referee whose comments and suggestions 
improved this work. We thank K. Milliman for sharing the H$\alpha$ BSS temperatures before publication. 
We are extremely grateful to P. Bergeron for sharing his grid of WD atmosphere models. N.~M.~G. 
acknowledges support from the W. J. McDonald Postdoctoral Fellowship.  R.~D.~M. 
and N.~M.~G. (while previously at UW-Madison) are supported through HST Program number 12492, provided by NASA 
through a grant from the Space Telescope Science Institute, which is operated by the Association of Universities 
for Research in Astronomy, Incorporated, under NASA contract NAS5-26555.  A.~S. is supported by the 
Natural Sciences and Engineering Research Council of Canada. A.~M.~G. is funded by a National Science 
Foundation Astronomy and Astrophysics Postdoctoral Fellowship under Award No. AST-1302765.


\begin{thebibliography}{}

\bibitem[Aarseth(2003)]
{nbody6} Aarseth, S. J.\ 2003, Gravitational N-Body Simulations (Cambridge: Cambridge Univ. Press)

\bibitem[Abate 
et al.(2013)]{Abate13} Abate, C., Pols, O.~R., Izzard, R.~G., 
Mohamed, S.~S., \& de Mink, S.~E.\ 2013, \aap, 552, A26 


\bibitem[Boffi 
et al.(2008)]{Boffi08} Boffi, F.~R., Sirianni, M., Lucas, R.~A.,
Walborn, N.~R., \& Proffitt, C.~R.\ 2008, Technical Instrument Report 
ACS 2008-002

\bibitem[Burbidge 
\& Sandage(1958)]{Burbidge58} Burbidge, E.~M., \& Sandage, A.\ 1958, \apj, 128, 174 

\bibitem[Carney 
et al.(2001)]{Carney01} Carney, B.~W., Latham, 
D.~W., Laird, J.~B., Grant, C.~E., \& Morse, J.~A.\ 2001, \aj, 122, 3419 

\bibitem[Castelli 
\& Kurucz(2004)]{atlas9} Castelli, F., \& Kurucz, R.~L.\ 2004, in IAU Symposium 210, 
Modelling of Stellar Atmospheres, ed. N. Piskunov et al., 
A20, arXiv:astro-ph/0405087 

\bibitem[Chen 
\& Han(2008)]{Chen08} Chen, X., \& Han, Z.\ 2008, \mnras, 387, 1416 

\bibitem[Dieball 
et al.(2005)]{Dieball05} Dieball, A., Knigge, 
C., Zurek, D.~R., et al.\ 2005, \apjl, 634, L105 

\bibitem[Feldman 
et al.(2010)]{Feldman10} Feldman, P.~D., Weaver, 
H.~A., Saur, J., \& McGrath, M.~A.\ 2010, Hubble after SM4.~Preparing JWST 

\bibitem[Ferraro
et al.(1995)]{Ferraro95} Ferraro, F.~R.,
Fusi Pecci, F., Bellazzini, M., et al.\ 1995, \aap, 294, 80

\bibitem[Ferraro 
et al.(1999)]{Ferraro99} Ferraro, F.~R., 
Paltrinieri, B., Rood, R.~T., \& Dorman, B.\ 1999, \apj, 522, 983 

\bibitem[Ferraro 
et al.(2001)]{Ferraro01} Ferraro, F.~R., 
D'Amico, N., Possenti, A., et al.\ 2001, \apj, 561, 337

\bibitem[Ferraro
et al.(2006)]{Ferraro06} Ferraro, F.~R., 
Sabbi, E., Gratton, R., et al.\ 2006, \apj, 647, 53

\bibitem[Ferraro 
et al.(2009)]{Ferraro09} Ferraro, F.~R., 
Beccari, G., Dalessandro, E., et al.\ 2009, \nat, 462, 1028 


\bibitem[Geller 
et al.(2013)]{Geller13} Geller, A.~M., Hurley, J.~R., \& Mathieu, R.~D.\ 2013, \aj, 145, 8 

\bibitem[Geller 
\& Mathieu(2012)]{Geller12} Geller, A.~M., \& Mathieu, R.~D.\ 2012, \aj, 144, 54 

\bibitem[Geller 
\& Mathieu(2011)]{Geller11} Geller, A.~M., \& Mathieu, R.~D.\ 2011, \nat, 478, 356 

\bibitem[Geller 
et al.(2009)]{Geller09} Geller, A.~M., Mathieu, 
R.~D., Harris, H.~C., \& McClure, R.~D.\ 2009, \aj, 137, 3743 

\bibitem[Geller 
et al.(2008)]{Geller08} Geller, A.~M., Mathieu, 
R.~D., Harris, H.~C., \& McClure, R.~D.\ 2008, \aj, 135, 2264 

\bibitem[Gonzaga 
et al.(2013)]{acshandbook} Gonzaga, S., et al., 2013, ACS Data Handbook, Version 7.1 (Baltimore: STScI).

\bibitem[Gosnell 
et al.(2014)]{Gosnell14} Gosnell, N.~M., 
Mathieu, R.~D., Geller, A.~M., et al.\ 2014, \apjl, 783, L8 

\bibitem[Henden 
\& Munari(2014)]{apass} Henden, A., \& Munari, U.\ 2014, 
Contributions of the Astronomical Observatory Skalnate Pleso, 43, 518 


\bibitem[Holberg 
\& Bergeron(2006)]{Holberg06} Holberg, J.~B., \& Bergeron, P.\ 2006, \aj, 132, 1221 

\bibitem[Hurley 
et al.(2002)]{Hurley02} Hurley, J.~R., Tout, C.~A., \& Pols, O.~R.\ 2002, \mnras, 329, 897 

\bibitem[Hurley 
et al.(2005)]{Hurley05} Hurley, J.~R., Pols, 
O.~R., Aarseth, S.~J., \& Tout, C.~A.\ 2005, \mnras, 363, 293 

\bibitem[Johnson 
\& Sandage(1955)]{Johnson55} Johnson, H.~L., \& Sandage, A.~R.\ 1955, \apj, 121, 616 

\bibitem[Knigge 
et al.(2009)]{Knigge09} Knigge, C., Leigh, N., 
\& Sills, A.\ 2009, \nat, 457, 288 

\bibitem[Knigge 
et al.(2008)]{Knigge08} Knigge, C., Dieball, A., 
Ma{\'{\i}}z Apell{\'a}niz, J., et al.\ 2008, \apj, 683, 1006 

\bibitem[Knigge 
et al.(2006)]{Knigge06} Knigge, C., Gilliland, 
R.~L., Dieball, A., et al.\ 2006, \apj, 641, 281 

\bibitem[Kroupa(2001)]
{Kroupa01} Kroupa, P.\ 2001, \mnras, 322, 231 

\bibitem[Leigh, Sills, 
\& Knigge(2007)]{Leigh07} Leigh, N., Sills, A., \& Knigge, C.\ 2007, \apj, 661, 210

\bibitem[Leigh 
\& Sills(2011)]{Leigh11} Leigh, N., \& Sills, A.\ 2011, \mnras, 410, 2370 

\bibitem[Leigh, Sills, 
\& Knigge(2011)]{Leigh11b} Leigh, N., Sills, A., \& Knigge, C.\ 2011, \mnras, 416, 1410

\bibitem[Leigh 
\& Geller(2012)]{Leigh12} Leigh, N., \& Geller, A.~M.\ 2012, \mnras, 425, 2369

\bibitem[Leigh 
et al.(2013)]{Leigh13} Leigh, N., Knigge, C. Sills, A., et al.\ 2013, \mnras, 428, 897

\bibitem[Leonard(1989)]
{Leonard89} Leonard, P.~J.~T.\ 1989, \aj,
98, 217

\bibitem[Leonard(1996)]
{Leonard96} Leonard, P.~J.~T.\ 1996, \apj, 
470, 521 

\bibitem[Lombardi 
et al.(2002)]{Lombardi02} Lombardi, J.~C., Jr., Warren, J.~S., Rasio, F.~A., Sills, A., 
\& Warren, A.~R.\ 2002, \apj, 568, 939 

\bibitem[Mapelli et 
al.(2006)]{Mapelli06} Mapelli, M., Sigurdsson, 
S., Ferraro, F.~R., et al.\ 2006, \mnras, 373, 361

\bibitem[Mapelli et 
al.(2007)]{Mapelli07} Mapelli, M., Ripamonti, 
E., Tolstoy, E., et al.\ 2007, \mnras, 380, 1127


\bibitem[Martin 
et al.(2005)]{galex} Martin, D.~C., Fanson, 
J., Schiminovich, D., et al.\ 2005, \apjl, 619, L1 

\bibitem[Mathieu 
\& Geller(2014)]{Mathieu14} Mathieu, R.~D., \& Geller, A.~M.\ 2014, in Ecology of Blue Stragglers, 
ed. H. Boffin, G. Carraro, \& G. Beccari, arXiv:1406.3467 

\bibitem[Mathieu 
\& Geller(2009)]{Mathieu09} Mathieu, R.~D., \& Geller, A.~M.\ 2009, \nat, 462, 1032 

\bibitem[Mathieu(2000)]
{Mathieu00} Mathieu, R.~D.\ 2000, in ASP Conf. Ser. 198, Stellar Clusters and 
Associations: Convection, Rotation, and Dynamos, ed. R. Pallavicini, G. Micela, \& S. Sciortino 
(San Francisco, CA: ASP), 517

\bibitem[McCrea(1964)]{McCrea64} McCrea, W.~H.\ 1964, \mnras, 128, 147

\bibitem[Meibom 
et al.(2009)]{Meibom09} Meibom, S., Grundahl, 
F., Clausen, J.~V., et al.\ 2009, \aj, 137, 5086 



\bibitem[Momany 
et al.(2007)]{Momany07} Momany, Y., Held, E.~V., Saviane, I., et al.\ 2007, \aap, 468, 973 

\bibitem[Naoz 
\& Fabrycky(2014)]{Naoz14} Naoz, S., \& Fabrycky, D.~C.\ 2014, \apj, 793, 137 

\bibitem[Paczy{\'n}ski(1971)]
{Paczynski71} Paczy{\'n}ski, B.\ 1971, \araa, 9, 183 

\bibitem[Perets 
\& Fabrycky(2009)]{Perets09} Perets, H.~B., \& Fabrycky, D.~C.\ 2009, \apj, 697, 1048 

\bibitem[Perets(2014)]{Perets14} Perets, H.~B.\ 2014, in Ecology of Blue Stragglers, 
ed. H. Boffin, G. Carraro, \& G. Beccari, arXiv:1406.3490 

\bibitem[Piotto et al.(2004)]{Piotto04} Piotto, G., De Angeli, F., King, I.,~R., et 
al.\ 2004, \apjl, 604, L109

\bibitem[Platais 
et al.(2003)]{Platais03} Platais, I., Kozhurina-Platais, V., Mathieu, R.~D., Girard, T.~M., 
\& van Altena, W.~F.\ 2003, \aj, 126, 2922 

\bibitem[Preston 
\& Sneden(2000)]{Preston00} Preston, G.~W., \& Sneden, C.\ 2000, \aj, 120, 1014 

\bibitem[Ram{\'{\i}}rez 
\& Mel{\'e}ndez(2005)]{Ramirez05} Ram{\'{\i}}rez, I., \& Mel{\'e}ndez, J.\ 2005, \apj, 626, 465 


\bibitem[Rodr{\'{\i}}guez-Merino 
et al.(2005)]{uvblue} 
Rodr{\'{\i}}guez-Merino, L.~H., Chavez, M., Bertone, E., \& Buzzoni, A.\ 2005, \apj, 626, 411 

\bibitem[Rozyczka 
et al.(2013)]{Rozyczka13} Rozyczka, M., Kaluzny, 
J., Thompson, I.~B., et al.\ 2013, AcA, 63, 67 

\bibitem[Rucinski(2000)]
{Rucinski00} Rucinski, S.~M.\ 2000, \aj, 
120, 319 


\bibitem[Sandage(1953)]
{Sandage53} Sandage, A.~R.\ 1953, \aj, 58, 
61 

\bibitem[Sandquist(2005)]
{Sandquist05} Sandquist, E.~L.\ 2005, \apjl, 635, 
L73

\bibitem[Santana et 
al.(2013)]{Santana13} Santana, F.~A., Munoz, R.~R., Geha, M., et al.\ 2013, \apj, 774, 106

\bibitem[Sarajedini 
et al.(1999)]{Sarajedini99} Sarajedini, A., von 
Hippel, T., Kozhurina-Platais, V., \& Demarque, P.\ 1999, \aj, 118, 2894 

\bibitem[Sills 
et al.(2001)]{Sills01} Sills, A., Faber, J.~A., 
Lombardi, J.~C., Jr., Rasio, F.~A., \& Warren, A.~R.\ 2001, \apj, 548, 323 

\bibitem[Sills 
et al.(2000)]{Sills00} Sills, A., Bailyn, C.~D., 
Edmonds, P.~D., \& Gilliland, R.~L.\ 2000, \apj, 535, 298 

\bibitem[Sills 
et al.(1997)]{Sills97} Sills, A., Lombardi, 
J.~C., Jr., Bailyn, C.~D., et al.\ 1997, \apj, 487, 290

\bibitem[Skrutskie 
et al.(2006)]{2mass} Skrutskie, M.~F., 
Cutri, R.~M., Stiening, R., et al.\ 2006, \aj, 131, 1163 

\bibitem[Talamantes 
et al.(2010)]{Talamantes10} Talamantes, A., Sandquist, E.~L., Clem, J.~L., et al.\ 2010, \aj, 140, 1268


\bibitem[Tian 
et al.(2006)]{Tian06} Tian, B., Deng, L., Han, Z., \& Zhang, X.~B.\ 2006, \aap, 455, 247 


\bibitem[Tremblay 
et al.(2011)]{Tremblay11} Tremblay, P.-E., 
Bergeron, P., \& Gianninas, A.\ 2011, \apj, 730, 128 


\bibitem[Webbink(1976)]
{Webbink76} Webbink, R.~F.\ 1976, \apj, 209, 829 

\bibitem[Wood(1995)]
{Wood95} Wood, M.~A.\ 1995, LNP Vol.~443: White Dwarfs, 41







\end{thebibliography}
\end{document}